# Centimeter-scale fully suspended metal and metal oxide thin films by one-step transfer-free liquid metal capillary forming

Chunlei Song[1], Zhenqi Guo[2], Yuanting Su[1], Changren Tian[1], Yeqi Zhu[1], Liang Lei[2], Jianbo Tang[*1,3]

[1]Department of Materials Science and Engineering, School of Engineering, Westlake University; Hangzhou 310030, China.

[2]Center for Advanced Engineering Sciences and Technology, School of Engineering, Westlake University; Hangzhou 310030, China.

[3]Research Center for Industries of the Future, Westlake University; Hangzhou 310030, China.

*Corresponding author. Email: jianbotang@westlake.edu.cn

Fully suspended thin films can decouple substrate effects and provide additional tuning degrees of freedom compared with their substrate-supported counterparts, making them unique platforms for next-generation thin film devices. Here we report one-step, transfer-free and substrate-free fabrication of centimeter-scale ultrathin fully suspended metal and metal oxide film structures *via* liquid metal capillary forming. We show that, analogous to soap film formation, the instantaneously developed few-nanometer-thick native surface oxide can laminate various liquid metals into micrometer-thick metallic films. Surprisingly, the surfactant-like metal oxide bilayer can survive dewetting-induced liquid metal drainage, forming suspended two-dimensional films featuring an enormous lateral size-to-thickness ratio on the order of $10^7$. We further demonstrate rapid prototyping of metallic minimal-surface thin-walled structures and ultra-sensitive acoustic wave detection with these suspended thin film platforms.

## Introduction

Metal and metal oxide thin films are essential in modern electronic, optical and energy systems[1–3]. Thin film technologies allow patternable material thin layers to be stacked and junctions to be created in a highly precise and integrated fashion. However, current thin film fabrication methods, such as physical and chemical vapor deposition[1], epitaxial growth[4], exfoliation[5–7], printing[8] and hot pressing[9,10], heavily rely on specialized substrates for supporting and/or directing thin film deposition and growth. The characteristics of the substrates, including material type, lattice structure, morphology and cleanness, in return affects the properties and quality of the resulted thin films. Suspended thin films[11], particularly in the two-dimensional (2D) regime, exhibit unconventional material behaviors in the absence of substrates and interfaces[12–17]. This unique material platform has enabled previously inaccessible probing techniques to be devised and novel devices to be developed for advancing thin film-based science and technology[18,19].

Despite the realization of various types of suspended thin films, they are typically grown on a substrate[15] or exfoliated from bulk crystals[17–19] before being transferred onto a suspending structure[20], that is, their fabrication still consists of one or more substrate-involving steps. Such a multistep substrate-involving growth-and-transfer strategy stringently restricts clean suspended thin films to be obtained[15] and, more importantly, their lateral dimensions to be substantially increased[17]. In addition, previously demonstrated ultra-thin suspended thin films are predominantly 2D materials (e.g., graphene, hexagonal boron nitride, transition metal dichalcogenide) deriving from layered crystals, while those from non-layered crystals (e.g., metal oxides) are rare.

In contrast to the lack of viable methods for fabricating suspended solid thin films with large lateral dimensions, suspended liquid films (for instance, soap films) form ubiquitously in nature as well as in numerous industrial processes[21,22]. To generate stable liquid films, capillary interactions must be balanced within a small thickness over large lateral dimensions, for which high surface tension and low viscosity are considered unfavorable[22–24]. For this reason, surfactants and thickeners (viscosity modifiers) are added to water in common practices to increase its film forming ability[25]. In this regard, liquid metals—metallic liquids that exhibit an enormous surface tension several times higher than that of water and water-like viscosity—are not expected to form stable thin films.

Counterintuitively, here we demonstrate that liquid metals can indeed form centimeter-scale fully suspended thin films with the assistance of their native surface oxide that develops spontaneously in ambient air. The metal oxide bilayer stabilizes the micrometer-thick *liquid metal thin films* (*l*-MTFs) sandwiched in between, in a way analogous to the familiar soap films stabilized by a surfactant molecule bilayer[23]. Unlike fleeting soap films, these *l*-MTFs can further be solidified to form everlasting thin-walled *solid metal thin films* (*s*-MTFs). Interestingly, the surface-developed metal oxide bilayer can further 'zip' together to form only a few-nanometer-thick suspended *metal oxide thin film* (MOTF) following the capillary drainage of the sandwiched liquid metal layer. Governed by capillary interactions which complete within a timescale of few seconds, these processes are able to generate centimeter-sized *l/s*-MTFs and MOTFs through the control of experimental parameters in a single, substrate-absent, transfer-free step. As such, the *liquid metal capillary forming* method reported in this work permits access to fully suspended forms of liquid/solid metal and metal oxide films as the playgrounds for future 2D material and thin film research.

## Realization of fully suspended MTFs and MOTFs

Fully suspended *l*-MTFs are fabricated in a way similar to soap film making (Fig. 1A). A ring-shaped copper (Cu) frame is immersed in a liquid metal bath of gallium (Ga, melting point, m.p.: 29.8 °C) kept at a constant temperature $T_B$. Due to the wetting of Cu by liquid Ga[26], withdrawing the frame from the bath causes the liquid metal to form a continuous film. Upon emerging above the bath, the Ga *l*-MTF instantaneously develops a self-limiting $GaO_x$ layer (referred to as a MOTF monolayer) on each side of the surface. The vertically lifted *l*-MTF is immediately positioned in a horizontal orientation to avoid rupture by gravity-driven flows. Analogous to the stabilizing effect provided by the viscoelastic surfactant bilayer to soap films[22,23,25], the MOTF bilayer laminates the high-surface tension liquid metal into a stable *l*-MTF (Fig. 1B). The *l*-MTFs could not be formed when we carried out control experiments in a nitrogen glovebox since the oxidation of the liquid metal is prohibited.

These *l*-MTFs exhibit several features that distinguish themselves from soap films (Fig. 1C). First, they are highly reflective (opaque) and not subject to evaporation or evaporation-induced instabilities and film rupture, giving rise to virtually everlasting, mirror-like liquid films. Second, the 'surfactant-like' metal oxide bilayer, being a viscoelastic solid phase[27], behaves differently compared to the surfactant molecule bilayer of soap films (Fig. 1C). One surprising consequence of having a metal oxide surfactant bilayer is the formation of a MOTF after the capillary drainage of the liquid metal layer (Fig. 2C). It is found that the liquid metal sandwiched between the two MOTF layers can be expelled to accumulate at the edge region of the frame *via* thin film dewetting (Fig. 1D) [28]. Meanwhile, the MOTF bilayer can be sustained in the absence of the inner liquid metal layer, forming a highly transparent 'zipped' bilayer MOTF (Fig. 1E). By contrast, the bilayer structure built by surfactant molecules in soap films will collapse instantaneously in the absence of the inner liquid layer.

As shown in Fig. 1F, the diameter of the ring-shaped frame ($D$), the diameter of its Cu wire ($d$), and the temperature of the liquid metal bath ($T_B$) collectively determine the formation of *l*-MTFs (regime I), its transition to MOTFs (regime II & III), and further to unstable regimes (regime IV & V). For a given $d = 0.6$ mm, small frame diameters (e.g., $D = 5$ mm) favor the formation of *l*-MTFs that are long-term stable even when small perturbations are intentionally applied (regime I). As $D$ progressively increases (e.g., $D = 10$ & 15 mm), the *l*-MTFs first enter a transition state which permits liquid metal dewetting under subtle perturbations (regime II) and then a spontaneous dewetting regime (regime III, e.g., $D \geq 20$ mm). Both regime II and regime III are accompanied by MOTF formation. Increasing $d$ expands the unstable regime V, indicating that thinner frames favor film formation—a trend also observed in soap films[29]. The temperature of the liquid metal bath influences the regime selection by affecting surface oxide formation and thermal stress built up in the oxide. High bath temperatures (e.g., $T_B \geq 150$ °C for Ga) lead to unstable regimes where a void appears and occupies partially (regime IV) or entirely (regime V) the thin films. Using a ternary bismuth-indium-tin alloy (BiInSn, m.p.: 62.0 °C), we reproduce the MTFs and MOTFs, as well as phase diagrams with similar parametric dependences (Fig. 1G). These results demonstrate that the liquid metal capillary forming method can be extended to other liquid metal and alloy systems.

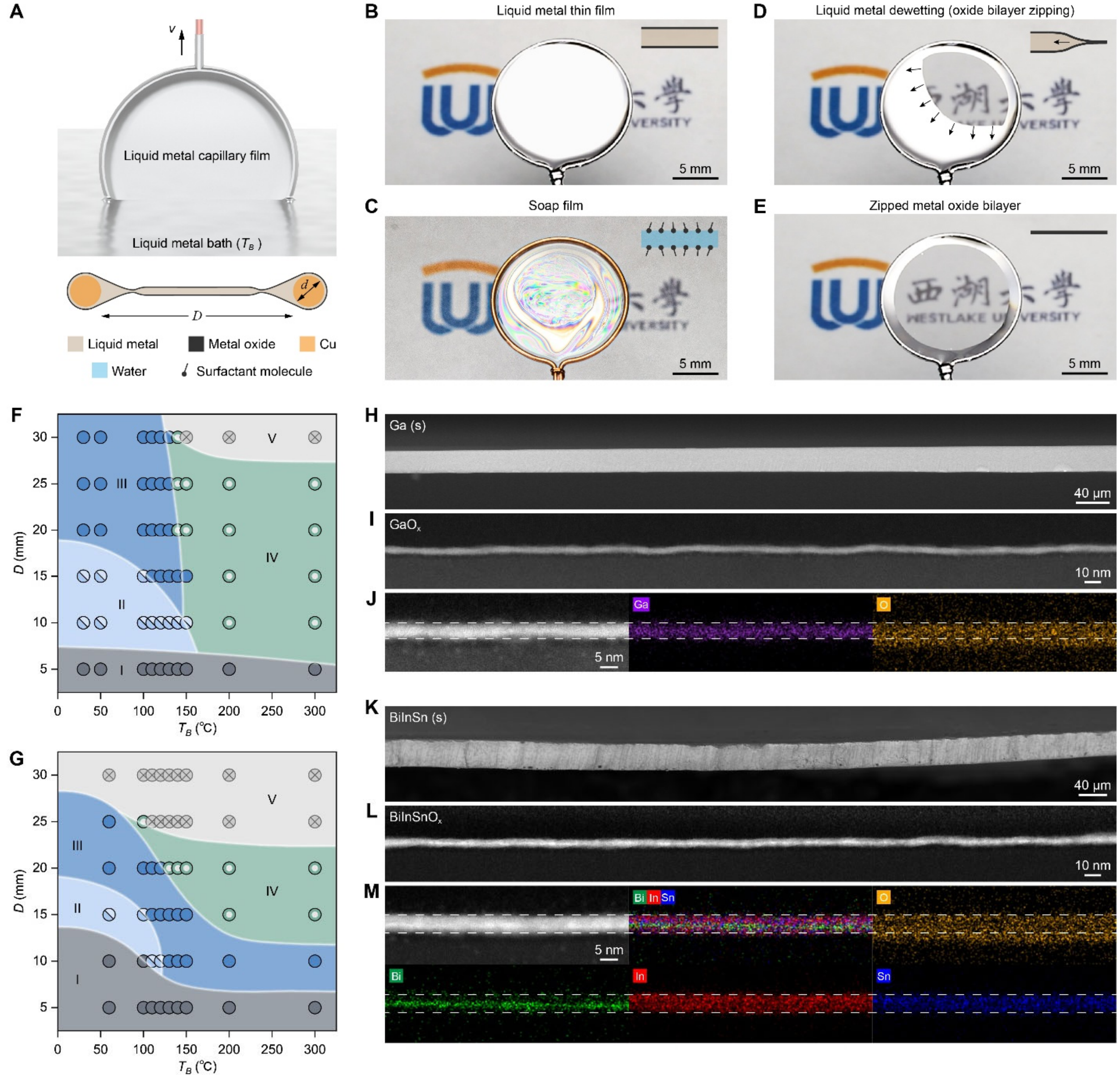

**Fig. 1. Formation and characterization of fully suspended MTFs and MOTFs.** **A**, Schematic experimental setup for fabricating *l*-MTFs suspended on a Cu frame (frame diameter *D* and wire diameter *d*). **B**-**E**, Photograph of a fully suspended Ga *l*-MTF (**B**, $D = 15$ mm and $d = 0.6$ mm), the same *l*-MTF during liquid metal dewetting with arrows indicating the liquid metal retreating direction (**D**), the fully-dewetted MOTF (**E**), and a suspended soap film (**C**). The schematics in the top-right corner of figures **B**-**E** depicts the cross-sectional configurations of the thin films. **F**,**G**, $T_B$–$D$ phase diagrams for Ga (**F**) and BiInSn (**G**) with $d = 0.6$ mm. **H**-**M**, Cross-sectional SEM images of a Ga *s*-MTF (**H**) and a BiInSn *s*-MTF (**K**). Cross-sectional HAADF-TEM images of a $GaO_x$ (**I**) and a $BiInSnO_x$ (**L**) film. Energy dispersive X-ray spectroscopy (EDS) maps of the 2D $GaO_x$ (**J**) and $BiInSnO_x$ (**M**) transferred onto a Si substrate. Due to the presence of a thin $SiO_x$ layer on the Si substrate, the O layer is slightly thicker than what is expected from the $GaO_x$ and $BiInSnO_x$ films.

Since the liquid metals are nonvolatile and have a mild m.p. above room temperature, their *l*-MTFs can be solidified into thin-walled solid metallic structures and be long-term maintained under ambient conditions. Scanning electron microscopy (SEM) of the cross section of the solidified Ga and BiInSn MTFs reveals that these films have an overall thickness of few tens of micrometers (Fig. 1H,K). Note that the MOTFs on the surfaces of the Ga and BiInSn MTFs are too thin to be observable using SEM. Examining the solidification microstructures of the fully suspended BiInSn alloy finds more organized solidification patterns than those formed in a BiInSn MTF of a similar thickness but under substrate-confined solidification. This highlights the substantial influence of the freely suspended state for mitigating the substrate influence during liquid-to-solid phase transition[30].

The fully suspended MOTFs resulting from liquid metal drainage are transferred onto a silicon (Si) substrate and then sliced by focused ion beam (FIB) for cross-sectional characterization using spherical aberration-corrected transmission electron microscopy (TEM). High-angle annular dark-field (HAADF) TEM images show that the $GaO_x$ (Fig. 1I,J) and $BiInSnO_x$ (Fig. 1L,M) thin film each has a consistent thickness of few nanometers, which agrees with the native oxide layer thickness of these liquid metals[31]. Cross-sectional elemental mappings reveal that while Ga distributes uniformly across the 2D $GaO_x$ (Fig. 1J), the constituent metals of $BiInSnO_x$ (Fig. 1M) display varied distributions, with Bi enriching the center region. This result suggests that Bi oxide tends to form in the vicinity of the liquid BiInSn alloy surface (prior to liquid metal dewetting), creating a concentration gradient along the few-nanometer thickness of the $BiInSnO_x$.

**Formation mechanism and thickness profile of the MTFs**

The thickness profile of a Ga *s*-MTF (Fig. 2A,B) and a BiInSn *s*-MTF is determined by micro-computed tomography (micro-CT) to gain mechanistic insight into *l*-MTF formation. Both MTFs consist of three distinct zones: i) a Plateau zone featuring a varying sub-millimeter-scale thickness, ii) a narrow minimum-thickness dimpled zone, and iii) a central uniform-thickness Frankel zone, which are signatures of soap films[22,32]. Such striking similarities in thickness profile imply that the formation of the *l*-MTFs and soap films shares a similar capillary origin, with the MOTF bilayer acting as the '*surfactant*' for the *l*-MTFs. Given that the characteristic thickness profile of capillary films is observed with the *s*-MTFs, it is expected that solidification-induced change to the overall thickness profile is minimal. Under the Reynolds lubrication approximation, the gradual establishment of the *l*-MTFs thickness profile can be described as[32],

$$\frac{\partial h}{\partial t} = -\frac{1}{r}\frac{\partial}{\partial r}\left[\frac{\sigma}{3\eta} r h^3 \frac{\partial}{\partial r}\left(\frac{1}{r}\frac{\partial}{\partial r}\left(r\frac{\partial h}{\partial r}\right)\right) + r v h\right] \tag{1}$$

where $h$ is the film thickness, $\sigma$ and $\eta$ the effective surface tension and viscosity of the liquid metal, $r$ the radial distance in a cylindrical coordinate, and $t$ is time. Solving this equation at different time steps shows the evolution of *l*-MTF thickness profile into the signature Plateau zone, dimpled zone and Frankel zone (Fig. 2C).

To demonstrate that the general applicability of this method to other liquid metals and alloys, fully suspended *l*-MTFs or *s*-MTFs are successfully fabricated from a gallium-indium alloy (GaIn, m.p. 15.5 °C), an indium-tin alloy (InSn, m.p. 120.0 °C), In (m.p. 156.6 °C), Sn (m.p. 231.9 °C), Bi (m.p. 271.5 °C), and aluminum (Al, m.p. 660.3 °C). The photographs and micro-CT images of the metallic thin films with varied m.p. are presented in Fig. 2D. The obtained *l*-MTFs (e.g., Ga & GaIn) are highly reflective and featureless, indicating that the liquid metal surface is exceptionally smooth and clean. The metals and alloys with a m.p. substantially higher than room temperature (e.g., BiInSn, InSn, In, Sn, Bi & Al) solidify into *s*-MTFs upon exiting the liquid metal bath, showing a less-reflective textured surface. The *s*-MTF of Bi, which is known for its interference color when forming thickness-varying surface oxide, display an iridescent appearance.

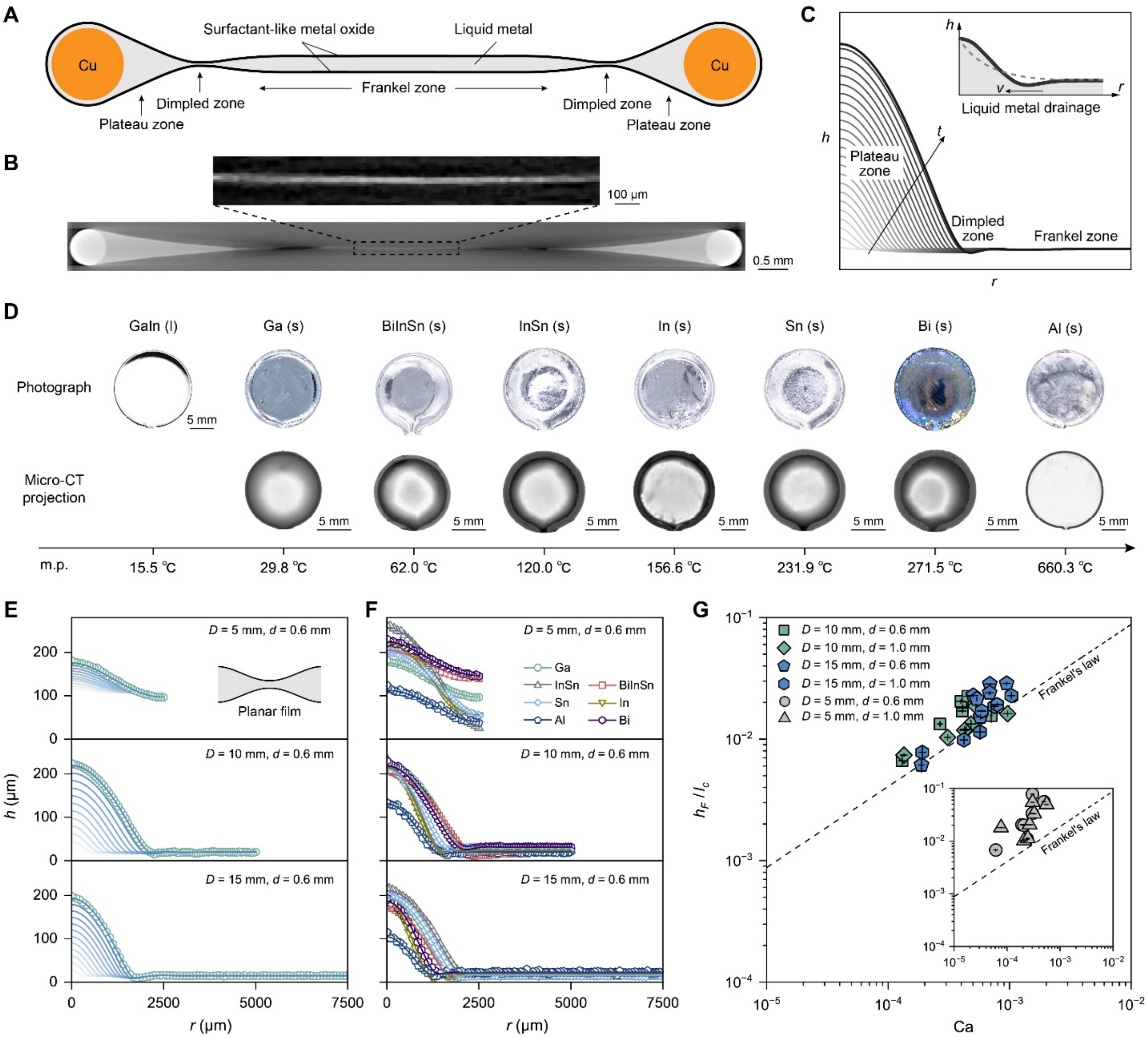

**Fig. 2. Formation mechanism and thickness profiles of MTFs. A**, Schematic cross-sectional thickness profile of a *l*-MTF. **B**, Cross-sectional micro-CT image of a Ga *s*-MTF with the Frankel zone magnified ($D$ = 10 mm and $d$ = 0.6 mm). **C**, Time-dependent thickness profile of a *l*-MTF. The inset schematically depicts the capillarity-driven thickness evolution of the *l*-MTF. **D**, Photographs and the corresponding grayscale micro-CT projections of different MTFs with their m.p. indicated. The GaIn *l*-MTF with a m.p. (15.5 °C) well below room temperature was not micro-CT scanned. **E**,**F**, Experimental (scatters) and calculated (lines) thickness profiles of Ga *s*-MTFs (**E**) and other metals and alloys examined (**F**) with different $D$ ($d$ = 0.6 mm). **G**, Normalized Frankel-zone thicknesses ($h_F/l_c$) of different MTFs as a function of Ca for $D$ = 10 and 15 mm. The inset shows the same plot for $D$ = 5 mm.

The micro-CT projections (Fig. 2D, bottom row) and the corresponding calculated cross-sectional profiles show that these MTFs develop three-zone thickness profile of typical capillary films (Fig. 2E,F). Exceptions are the small-diameter MTFs (e.g., $D$ = 5 mm), where the Plateau zones extend to the center of the films. Both $D$ and $d$ are found to modify the thickness profile of the $l$-MTFs. In the absence of the Frankel zone and the dimpled zone, these planar films show substantially increased film thickness with the minimum found at their center. Recalling the phase diagrams of the $l$-MTFs (Fig. 1F,G), the stability of the $l$-MTFs in regime I at small $D$ is expected from the fact that these thick planar films are dimple free[23]. At increased $D$ (regime II and III), the minimum-thickness dimpled zone becomes a *weak point* of the $l$-MTFs at which dewetting always initiates. Though the dimpled zone destabilizes the $l$-MTFs, its dewetting susceptivity provides the key mechanism for generating fully suspended MOTFs.

The film thickness of the Frankel zone ($h_F$) is calculated following Frankel's law[33,34]: $h_F = kl_c \, \mathrm{Ca}^{2/3}$, where $l_c = (\sigma/\rho g)^{1/2}$ is the capillary length, $\rho$ the density of the liquid metal, $g$ the gravitational acceleration, the coefficient $k = 1.89$[33], and $\mathrm{Ca} = \eta v/\sigma$ the capillary number. As shown in Fig. 2G, the normalized thickness ($h_F/l_c$) of different MTFs can be well fitted by Frankel's law for $D$ = 10 mm and 15 mm. When the Frankel zone is absent ($D$ = 5 mm), deviations of the film thickness from Frankel's law are observed (Fig. 2G, inset). These mechanistic insights prove that the $l$-MTFs are capillary films having well-defined thickness profiles.

## Formation and characterization of MOTFs

Provided that the solidification of a liquid metal can be prohibited (e.g., Ga & GaIn) or delayed (by supercooling, e.g., BiInSn), the $l$-MTFs transform into fully suspended MOTFs *via* liquid metal dewetting. The method is able to fabricate large-area fully suspended MOTFs with different shapes (Fig. 3A). Particularly, circular $GaO_x$ films with a diameter up to 2.5 cm (1-inch wafer size) and rectangular $GaO_x$ films with one edge extending to as long as 4 cm (~4 $cm^2$ in area) are obtained, leading to an enormous lateral size-to-thickness ratio on the order of $10^7$. Furthermore, these MOTFs can be readily transferred onto various substrates, including hollow grids with a high success rate (Fig. 3B). These wafer-sized, few-nanometer-thick, fully suspended MOTFs exceeds previously-reported suspended 2D films by at least one order of magnitude, by either suspended area or area-to-thickness ratio (Fig. 3C).

The obtained 2D $GaO_x$ (Fig. 3D-F) and $BiInSnO_x$ (Fig. 3G-I) thin films are amorphous[35], observing from both their in-plane (Fig. 3E,H) and cross-sectional (Fig. 3F,I) directions by high-resolution TEM. The amorphous nature of these liquid metal-derivate oxides is further evidenced by the selected-area electron diffraction patterns (insets of Fig. 3E,H). Comparatively, $BiInSnO_x$ shows more pronounced short-range order than $GaO_x$ (Fig. 3E,H). In addition, these MOTFs are uniformly spread with liquid metal nanodroplets which stem from the dewetting instabilities under the confinement of the metal oxide bilayer. Atomic force microscopy (AFM) reveals that the thicknesses of the zipped MOTF bilayers are not simply twice the thickness of the surface-exfoliated MOTF monolayers. Instead, the zipped bilayers show a considerably reduced thickness (Fig. 3J-M). The fact that the TEM cross sections of the zipped MOTF bilayers display no distinguishable interface (Fig. 3F,I) suggests an atomic zipping mechanism for the anomalous thickness decrease of the MOTFs, that is, inter-layer chemical bonds zip the bilayer together upon the two monolayers come into contact, causing the MOTF thinning by flattening out subtle surface irregularities.

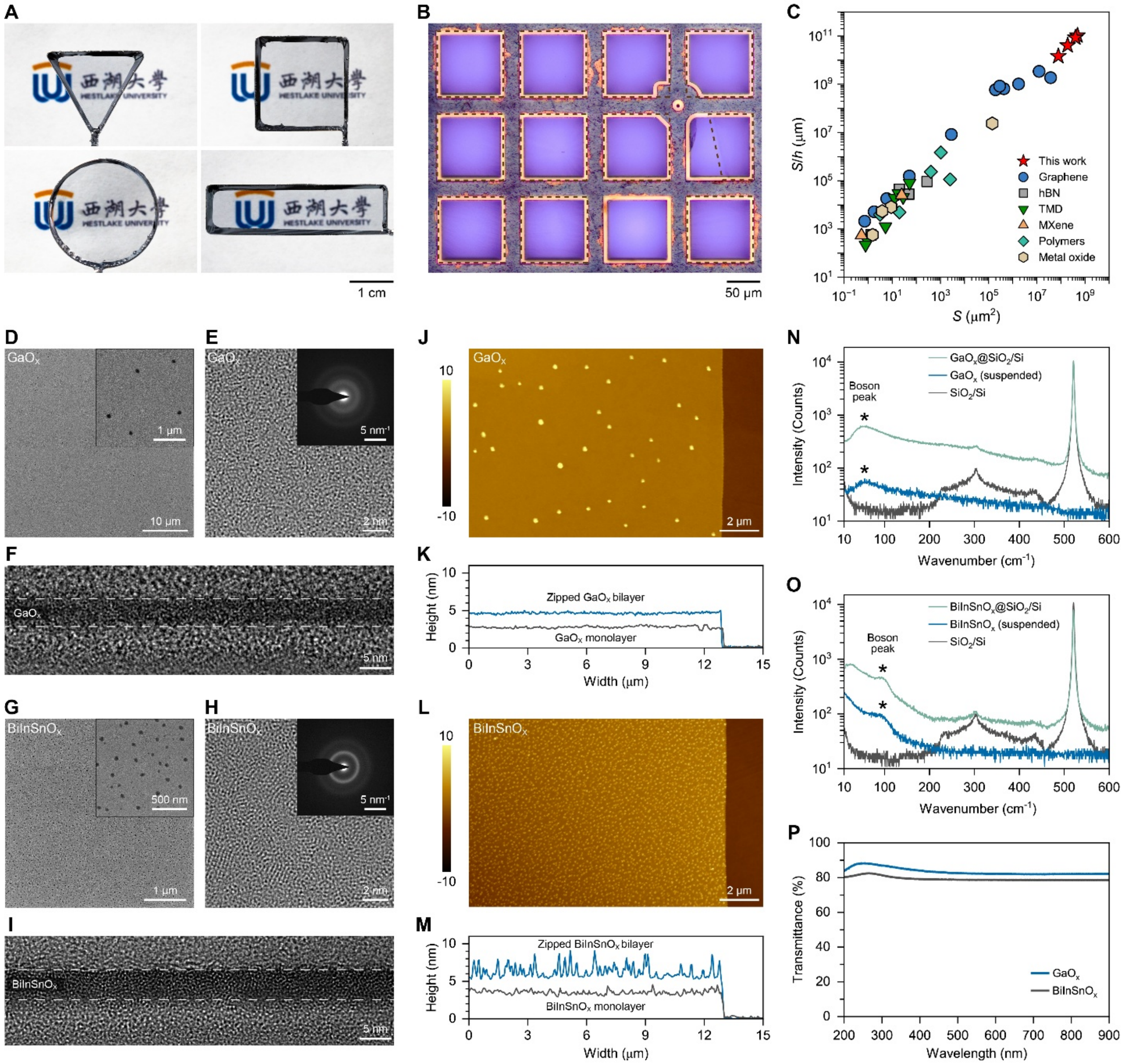

**Fig. 3. Characterization of suspended MOTFs.** **A**, Photographs of $GaO_x$ films suspended on Cu frames with various sizes and shapes. **B**, Suspended $GaO_x$ film transferred onto a carbon-layer-removed Cu TEM grid (dashed boxes indicate the regions covered with suspended $GaO_x$). **C**, Comparison of the area *S* and the area-to-thickness ratio *S/h* of fully suspended 2D films. **D-I**, TEM images of $GaO_x$ (**D-F**) and $BiInSnO_x$ (**G-I**) films suspended on a Cu TEM grid (carbon coating layer removed). Low-resolution TEM image with the inset showing liquid metal nanodroplet inclusion in the MOTF (**D**,**G**). High-resolution TEM image with the inset showing the corresponding selected-area electron diffraction pattern (**E**,**H**), and the cross-sectional TEM image (**F**,**I**). **J-M**, AFM topography of a zipped MOTF bilayer of $GaO_x$ (**J**) and $BiInSnO_x$ (**L**), along with the thickness profiles of the zipped MOTF bilayer compared with those of a monolayer of the same MOTFs (**K**,**M**). **N**,**O**, Raman spectra of the suspended $GaO_x$ (**N**) and $BiInSnO_x$ (**O**) films compared against that of the substrate and substrate-supported films. The asterisks indicate the boson peaks. **P**, Optical transmittance spectra of the fully suspended $GaO_x$ and $BiInSnO_x$ films.

Low-wavenumber Raman spectroscopy reveals a broad peak at 56.6 $cm^{-1}$ for the $GaO_x$ film (Fig. 3N) and 96.4 $cm^{-1}$ for the $BiInSnO_x$ film (Fig. 3O). These are boson peaks indicative of the amorphous nature of these MOTFs[36]. To the best of our knowledge, this is also the first identification of boson peaks in liquid metal-derivate 2D metal oxides. In comparison with the substrate-supported samples, the suspended MOTFs display a single intensity-reduced yet salient boson peak, implying that suspended MOTFs can avoid potential interfering signals arising from the substrate. Owing to their ultra-thin nature, both the $GaO_x$ and $BiInSnO_x$ films exhibit a high transmittance of ~80% or greater across a broad electromagnetic wavelength range in our optical transmittance spectroscopy measurement (Fig. 3P). Notably, both MOTFs show a near-constant transmittance in the visible and the following near infrared spectrum.

## Rapid prototyping of thin-walled 3D structures

By designing suitable frame geometries to define the boundaries of the *l*-MTFs, typical minimal surfaces observed with soap films[21] are reproduced with liquid metal capillary films. Figure 4A demonstrates a Ga *s*-MTF catenoid suspended between two parallel ring-shaped frames with a circular $GaO_x$ thin film in the catenoid center. The classic helicoid thin-film geometry is successfully fabricated using a circular helix frame and the InSn alloy as the liquid metal (Fig. 4B). Furthermore, using a fullerene-shape (or a buckyball-shape) frame, a miniature metallic *soccer ball* with its pentagonal and hexagonal facets enclosed by Ga *s*-MTFs is obtained (Fig. 4C).

As can be observed from their cross-sectional micro-CT images, the influence of gravity is more pronounced for these 3D *l*-MTFs, pulling more liquid metal to the lower side and increasing the film thickness therein. In addition, capillary bridges are clearly seen in between the frame boundaries. It should be note that these 3D objects, featuring tens-of-micrometer thicknesses and complex geometries, are fabricated within a timescale of few seconds and without the use of any substrate or mold, which cannot be achieved with conventional methods (such as 3D printing and hot pressing/molding) for forming such thin-walled metallic structures. Given that various metals are processable (Fig. 2D), it can be a promising alternative for rapid prototyping of miniature (capillary scale) thin-walled 3D objects.

## MOTF-based acoustic wave detection

The vibrational responses of the fully suspended MOTFs are investigated to analyses their material properties and, as a proof-of-concept, employ them as ultra-sensitive diaphragms for acoustic wave detection. A customized experimental platform that comprises a tilted 2D $GaO_x$ film is built (Fig. 4D). This allows the horizontal projection of film vibration to be monitored using high-speed optical microscopy, by tracing the horizontally-projected trajectory of individual liquid metal nanodroplet inclusion[37] (Fig. 4E).The detected sinusoidal signals indicate that the vibration of the circular $GaO_x$ thin films is robust, sensitive, and predominantly in the (0,1) mode (Fig. 4E-G). The corresponding resonant frequency ($f_{0,1}$) satisfies[38]: $f_{0,1} = \lambda_{m,n}(2\pi a)^{-1}(\tau^*/\rho^*)^{1/2}$, where $\lambda_{m,n}$ is a vibration mode-dependent constant and $\lambda_{0,1} = 4.808$, $a$ the radius (slightly smaller than $D/2$), $\tau^*$ the tension per unit length, $\rho^*$ the density per unit area of the film. With $f_{0,1}$ and $\rho^*$ experimentally determined, the material-specific properties $\tau^*/\rho^*$ and $\tau^*$ of the 2D $GaO_x$ films are determined to be $59.76 \pm 11.20$ $m^2$ $s^{-2}$ and $4.60 \pm 0.90$ mN $m^{-1}$, respectively.

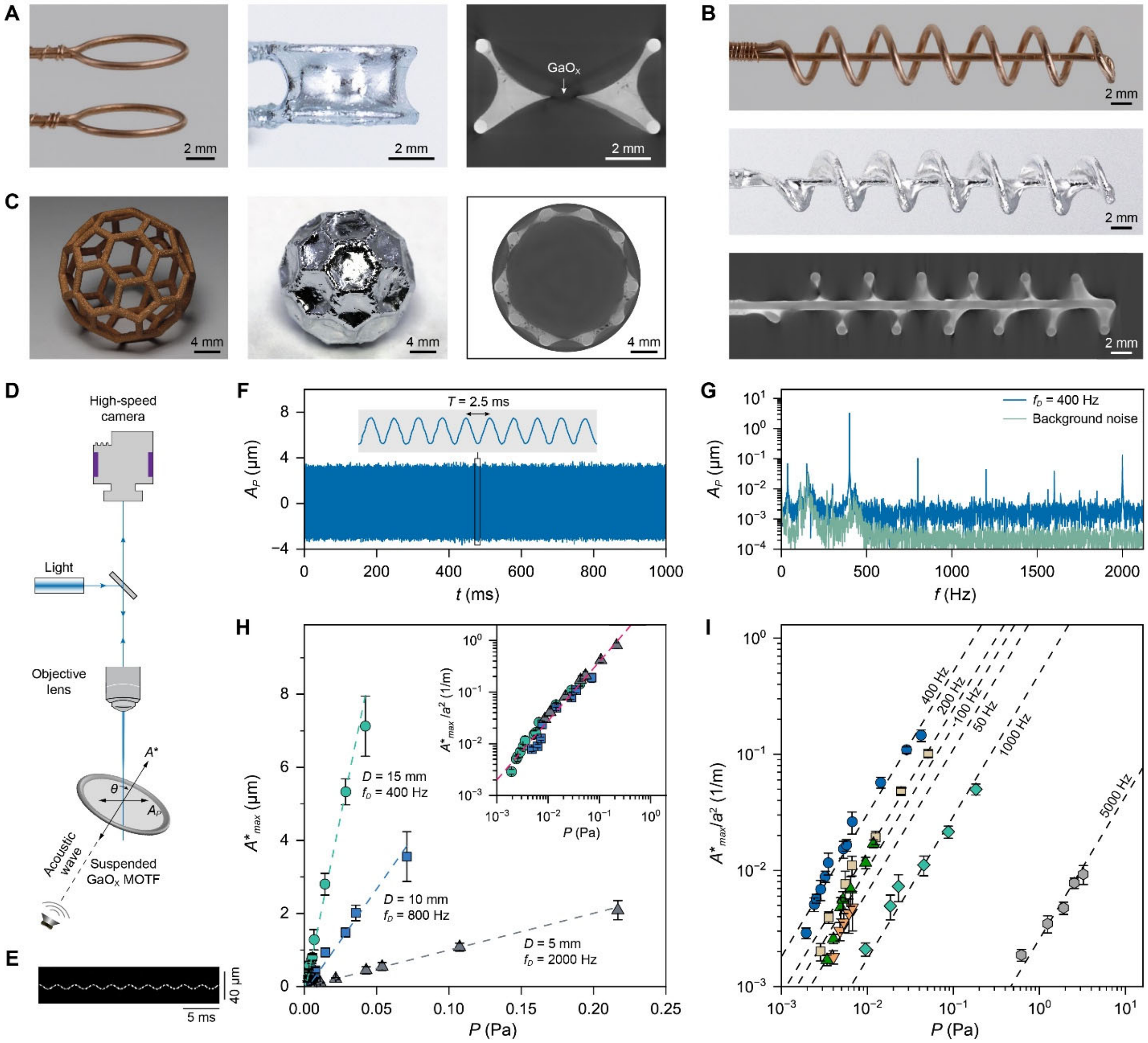

**Fig. 4. Demonstration of rapid prototyping of thin-walled 3D metallic objects and ultra-sensitive acoustic wave detection by MOTFs. A-C**, Thin-walled Ga catenoid (**A**), InSn helicoid (**B**) and Ga buckyball (*soccer ball*, **C**) fabricated by liquid metal capillary forming. A photograph of the Cu frame, the obtained MTF structures, and a cross-sectional micro-CT image are included in each figure set. **D**, Schematic experimental setup of the vibration detection platform. **E**, Sinusoidal vibration trajectory of a tracer reconstructed from the high-speed microscopy recording. **F**, Time-domain maximum vibration amplitude $A^*_{max}$ of a $GaO_x$ film ($D$ = 15 mm and $d$ = 0.6 mm) under 400 Hz harmonic excitation ($P$ = 0.042 Pa). **G**, Fourier transformation of (**F**) with multiple high-order vibration frequencies visible. The background signal is plotted for comparison. **H**, $A^*_{max}$-$P$ plot for $GaO_x$ films of different diameters driven at their respective near-resonate frequency. For $D$ = 5, 10 and 15 mm ($d$ = 0.6 mm fixed), $f_{0,1}$ is measured to be 1840, 744 and 380 Hz, respectively. The inset shows the normalized vibration amplitude $A^*_{max}/a^2$ of the same data sets. **I**, $A^*_{max}/a^2$ of $GaO_x$ films under different driving frequencies ($D$ = 15 mm and $d$ = 0.6 mm).

The maximum vibration amplitude $A^*_{max}$, measured at the film center scales linearly with the acoustic pressure $P$ in the millipascal to pascal range (Fig. 4H,I). Driven by a fixed near-resonate frequency, the slope of the vibration response curves increases with $D$ (Fig. 4H), meaning that the detection sensitivity scales with lateral size ($D$) of the films. The normalized amplitudes $A^*_{max}/a^2$ for different $GaO_x$ films fall onto a single linear curve (inset of Fig. 4H), indicating the linearly scalable influence of damping[39]. A similar linear relationship between $A^*_{max}/a^2$ and $P$ is also observed for the 2D $GaO_x$ films with $D = 15$ mm and $d = 0.6$ mm (Fig. 4I). As the driving frequency moves away from the resonate frequency ($f_{0,1}$ = 380 Hz), $A^*_{max}/a^2$ progressively decreases in the double logarithmic plot while their slope remains constant (Fig. 4I). According to these findings, the detection limit of vibration amplitude and acoustic pressure reach 80 nm and 2 mPa, respectively. The film with the largest size tested ($D = 15$ mm and $d = 0.6$ mm) exhibits an acoustic wave detection sensitivity of $1.94 \times 10^{-4}$ m/Pa, which is several orders of magnitude higher than that obtained with other 2D films and comparable to those based on thick membranes. The fully suspended MOTFs are able to provide vibrational detection with such high sensitivities since they combine exceedingly large (centimeter scale) sizes and extremely small thickness (few nanometers).

## Conclusion

We demonstrated that centimeter-scale fully suspended thin films of liquid metal, solid metal and metal oxide can be fabricated using a one-step, substrate-absent and transfer-free liquid metal capillary forming process. Mechanistic investigations and thin film characterizations show that a number of analogies can be made between the *l*-MTFs and soap films. However, the two are markedly different in material properties. These differences are responsible for the formation of the *s*-MTF structures and the MOTFs. The fabrication process is fast, allowing complex 3D thin-walled minimal surface objects and MOTFs to be formed in the timescale of few seconds. Being able to access fully suspended MTFs and MOTFs while completely avoiding substrate during both fabrication and use is highly desirable for achieving high film cleanness and avoiding substrate/interface effects. In achieving ultra-sensitive MOTF-based vibrational detection, we demonstrate the potential of the suspended MOTFs for enabling device capabilities that are unattainable previously. The proposed liquid metal capillary forming method is expected to provide a general strategy for realizing fully suspended metal and metal oxide thin films for next-generation thin film-based science and technology.

**Acknowledgments:** The TEM, AFM, Raman, UV-Vis, XPS characterizations of this project were carried out in the Instrumentation and Service Center for Physical Sciences and the Instrumentation and Service Center for Molecular Sciences at Westlake University. **Funding:** National Natural Science Foundation of China grant 92580116 (JT); National Natural Science Foundation of China grant 52571232 (JT); National Natural Science Foundation of China grant 12502315 (CS); Westlake University Research Center for Industries of the Future; Westlake Education Foundation.

**Author contributions:** Conceptualization: JT; Investigation: CS, ZG, YS, CT, YZ; Visualization: CS, JT; Software: CS, ZG; Funding acquisition: JT, CS; Project administration: JT, CS; Resource: JT, LL; Supervision: JT, LL; Writing – original draft: JT, CS; Writing – review & editing: JT, CS.